\documentstyle[aps,epsf]{revtex}

\begin{document}
\title{Gravitational collapse of a Hagedorn fluid in Vaidya geometry}
\author{T. Harko\thanks{%
E-mail: harko@hkucc.hku.hk}}
\address{Department of Physics, The University of Hong Kong, Pokfulam Road, Hong\\
Kong, P. R. China.}
\maketitle

\begin{abstract}
The gravitational collapse of a high-density null charged matter fluid,
satisfying the Hagedorn equation of state, is considered in the framework of
the Vaidya geometry. The general solution of the gravitational field
equations can be obtained in an exact parametric form. The conditions for
the formation of a naked singularity, as a result of the collapse of the compact
object, are also investigated. For an appropriate choice of the arbitrary
integration functions the null radial outgoing geodesic, originating from
the shell focussing central singularity, admits one or more positive roots.
Hence a collapsing Hagedorn fluid could end either as a black hole, or as a
naked singularity. A possible astrophysical application of the model, to
describe the energy source of gamma-ray bursts, is also considered.

\end{abstract}


\section{Introduction}

Investigating the final fate of the gravitational collapse of an initially
regular distribution of matter, in the framework of the Einstein theory of
gravitation, is one of the most active fields of research in contemporary
general relativity. One would like to know whether, and under what initial
conditions, gravitational collapse results in black hole formation. One
would also like to know if there are physical collapse solutions that lead
to naked singularities. If found, such solutions would be counterexamples of
the cosmic censorship hypothesis, which states that curvature singularities
in asymptotically flat space-times are always shrouded by event horizons.

Roger Penrose \cite{Pe69} was the first to propose the idea, known as cosmic
conjecture: does there exist a cosmic censor who forbids the occurrence of
naked singularities, clothing each one in an absolute event horizon? This
conjecture can be formulated in a strong sense (in a reasonable space-time
we cannot have a naked singularity) or in a weak sense (even if such
singularities occur they are safely hidden behind an event horizon, and
therefore cannot communicate with far-away observers). Since Penrose' s
proposal, there have been various attempts to prove the conjecture (see \cite
{Jo93} and references therein). Unfortunately, no such attempts have been
successful so far.

Since, due to the complexity of the full Einstein equations, the general
problem appears intractable, metrics with special symmetries are used to
construct gravitational collapse models. One such case is the
two-dimensional reduction of general relativity obtained by imposing
spherical symmetry. Even with this reduction, however, very few
inhomogeneous exact nonstatic solutions have been found. One well-known
example is the Vaidya metric \cite{Va51}. It describes the gravitational
field associated with the eikonal approximation of an isotropic flow of
unpolarized radiation, or, in other words, it represents a null fluid. It is
asymptotically flat and it is employed in modeling the external field of 
radiating stars and evaporating black holes. The second one is the
Tolman-Bondi metric \cite{To34}, which gives the gravitational field
associated with dust matter and is frequently applied either in cosmological
models or in describing the collapse of a star to a black-hole. Tolman-Bondi
space-times embody the Schwarzschild solution, the Friedman universes and
the Oppenheimer-Snyder collapse, as well as inhomogeneous expansions and
collapses.

At first sight these two metrics are completely different. Do the naked
singularities that form in the collapse of null radiation and in the
collapse of dust bear any relation with each other? Are there any features
common to both solutions? And if this is the case, what are the implications
for cosmic censorship? As shown by Lemos \cite{Le92}, the naked
singularities which appear in Vaidya and Tolman-Bondi space-times are of the
same nature. Various important features such as the degree of inhomogeneity
of the collapse necessary to produce a naked singularity, the Cauchy horizon
equation, the apparent horizon equation, the strength of the singularity and
the stability of the space-time have a mutual correspondence in both
metrics. For cosmic censorship, this result implies that if the
shell-focusing singularities arising from the collapse of a null fluid are
not artifacts of some (eikonal) approximation, then the shell-focusing
singularities arising from the collapse of dust are also not artifacts (and
vice versa). Conversely, if the naked singularities are artifacts in one of
them so are they in the other.

Thus the Vaidya metric belongs to the Tolman-Bondi family. The most unbound
case yields the Vaydia metric. Hence one expects that major features which
might arise in one of the metrics will also appear in the other. One example
is the result that the strength in the Vaidya metric depends on the
direction from which the geodesics enter the singularity \cite{Le92}.

Null fluids, are, in principle, easier to treat than matter fields. A null
fluid is the eikonal approximation of a massless scalar field. Thus if one
shows that the naked singularities arising in the Vaidya metric can be
derived from more fundamental (massless) fields, then the naked
singularities, which form in the Tolman-Bondi collapse, may also be derived
from more fundamental (massive) fields. The same types of relations and
conclusions hold for charged radiation and charged dust matter. The
structure and properties of singularities in the gravitational collapse in
Vaidya space-times have been analyzed, from different points of view,
in \cite{BoVa70}-\cite{Or98}.

Within the framework of various physical models, the spherical gravitational
collapse has been analyzed in many papers. The role of initial density and
velocity distributions towards determining the final outcome of spherical
dust collapse and the causal structure of the singularity has been examined
in terms of evolution of apparent horizon in \cite{SiJo96}, \cite{JhJoSi96}.
The collapse was described by the Tolman-Bondi metric with two free
functions. The collapse can end in either a black hole or a naked
singularity. The occurrence and nature of naked singularities in the
Szekeres space-times have been investigated in \cite{JoKr96}. These
space-times represent irrotational dust. They do not have any Killing
vectors and they are the generalizations of the Tolman-Bondi space-times.
There also exist naked singularities that satisfy both the limiting focusing
condition and the strong limiting focusing conditions. The relevance of the
initial state of a collapsing dust cloud towards determining its final fate
in the course of a continuing gravitational collapse has been considered in 
\cite{DwJo97}. Given any arbitrary matter distribution for the cloud in the
initial epoch, there is always the freedom to chose the rest of the initial
data, namely the initial velocities of the collapsing spherical shells, so
that, depending on this choice, the collapse could result either in a black
hole or a naked singularity. Thus, given the initial density profile, to
achieve the desired end state of the gravitational collapse one has to give
a suitable initial velocity of the cloud. The expression for the expansion
of outgoing null geodesics in spherical dust collapse has been derived in 
\cite{Si97} The limiting values of the expansion in the approach to
singularity formation have been computed. Using these results one can show
that the covered, as well as the naked singularity solutions arising in
spherical dust collapse, are stable under small changes in the equation of
state.

The growth of the Weyl curvature is examined in two examples of naked
singularity formation in spherical gravitational collapse-dust and Vaidya
space-time- in \cite{BaSi97}. The Weyl scalar diverges along outgoing radial
null geodesics as they meet the naked singularity in the past. Although
general relativity admits naked singularities arising from gravitational
collapse, the second law of thermodynamics could forbid their occurrence in
nature. A simple model for a corona of a neutrino-radiating star showing
critical behavior is presented in \cite{Go98}. The conditions for the
existence or absence of a bounce (explosion) are discussed. The charged
Vaidya metric was extended to cover all of the space-time in \cite{PaWi98} and
the Penrose diagram for the formation and evaporation of a charged black
hole obtained. The covariant equations characterizing the strength of a
singularity in spherical symmetry and a slight modification to the
definition of singularity strength have been derived in \cite{No99}. The
idea of probing naked space-times singularities with waves rather than with
particles has been proposed in \cite{IsHo99}. For some space-times the
classical singularity becomes regular if probed with waves, while stronger
classical singularities remain singular.

In order to obtain the energy-momentum tensor for the collapse of a null
fluid an inverted approach was proposed by Husain \cite{Hu96}. First the
stress-energy momentum tensor is determined from the metric. Then the
equation of state and the dominant energy condition are imposed on its
eigenvalues. This leads to an equation for the metric function. The precise
form of the stress-energy tensor is then displayed. By using this approach
several classes of solutions describing the collapse of a null fluid,
satisfying barotropic and polytropic type equations of state, have been
obtained. In the framework of the same approach a large class of solutions,
including Type II fluids, and which includes most of the known solutions of
the Einstein field equations, has been derived, in four dimensions by Wang and
Wu \cite{WaWu98}, and in $N\geq 4$ dimensions by Villas da Rocha \cite{Ro02}.
The Vaidya radiating metric has been extended to include both a radiation
field and a string fluid by Glass and Krisch \cite{GlKr98}, \cite{GlKr99}
and by Govinder and Govender \cite{GoGo03}.

When nuclear matter is squeezed to a sufficiently high density, a phase
transition takes place and neutron matter converts into three-flavor
(strange) quark matter, which is due to the fact that strange matter may be
more stable than nuclear matter. The collapse of the quark fluid, described
by the bag model equation of state $p=\left( \rho -4B\right) /3$, with $B=$%
constant, has been studied by Harko and Cheng \cite{HaCh00} and the
conditions for the formation of a naked singularity have been obtained. The
obtained solution has been generalized to arbitrary space-time dimensions
and to a more general linear equation of state by Ghosh and Dadhich \cite
{GhDh02}, \cite{GhDh03}.

In 1965 Hagedorn \cite{Ha65} postulated that for large masses $m$ the
spectrum of hadrons $\rho \left( m\right) $ grows exponentially, $\rho
\left( m\right) \sim \exp \left( m/T_{H}\right) $, where $T_{H}$, the
Hagedorn temperature, is a scale parameter. The hypothesis was based on the
observation that at some point a further increase of energy in proton-proton
and proton-antiproton collisions no longer raises the temperature of the
formed fireball, but results in more and more particles being produced. Thus
there is a maximum temperature $T_{H}$ that a hadronic system can achieve.
The statistical model of the hadrons has been used to obtain a description
of dense matter at densities exceeding nuclear density. The Hagedorn phase
also arises in theories containing fundamental strings, because they have a
large number of internal degrees of freedom \cite{Gi89}. As a result of the
existence of many oscillator modes the density of states grows exponentially
with single string energy. Thermodynamical quantities, such as the entropy,
diverge at the Hagedorn temperature. If one considers an ensemble of weekly
interacting strings at finite temperature, this behavior of the density of
states is thought to lead either to a limiting temperature or a phase
transition, in which the string configuration changes to one which is
dominated by a single long string \cite{Gr01}. The high density Hagedorn
phase of matter has been extensively used in cosmology to describe the very
early phases of the evolution of the Universe \cite{Ma98}-\cite{Ba03}.

It is the purpose of the present paper to study the spherically symmetric
gravitational collapse of the charged matter in the Hagedorn phase. In order
to simplify the mathematical formalism we adopt the assumption that the high
density fluid moves along the null geodesics of a Vaidya type space-time.
The Vaidya geometry, also permitting the incorporation of the effects of the
radiation, offers a more realistic background than static geometries, where
all back reaction is ignored. By adopting the Hagedorn equation of state for
dense matter, the general solution of the field equations can be obtained in
an exact form. For the sake of comparison we also consider the collapse of
matter described by the stiff Zeldovich equation of state. The conditions of
formation of naked singularities are obtained in both cases.

The present paper is organized as follows. The field equations describing
the collapse of a Hagedorn fluid are written down in Section II. The general
solution of the field equations is presented in Section III. The equation of
the null outgoing geodesics and the conditions of the formation of the naked
singularities are discussed in Section IV. An astrophysical application of
the formalism to explain the gamma ray bursts energy emissions is described
in Section V. In Section VI we discuss and conclude our results.

\section{Geometry and field equations}

In ingoing Bondi coordinates $(u,r,\theta ,\varphi )$ and with advanced
Eddington time coordinate $u=t+r$ (with $r\geq 0$ the radial coordinate and $%
r$ decreasing towards the future) the line element describing the radial
collapse of a coherent stream of matter can be represented in the form \cite
{Hu96}, \cite{HaCh00} 
\begin{equation}
ds^{2}=-\left[ 1-\frac{2m\left( u,r\right) }{r}\right] du^{2}+2dudr+r^{2}%
\left( d\theta ^{2}+\sin ^{2}\theta d\varphi ^{2}\right) .  \label{222}
\end{equation}

$m\left( u,r\right) $ is the mass function and gives the gravitational mass
within a given radius $r$. In the following we use the natural system of
units with $8\pi G=c=1$.

The matter energy-momentum tensor can be written in the form \cite{Hu96}, 
\cite{WaWu98} 
\begin{equation}
T_{\mu \nu }=T_{\mu \nu }^{(n)}+T_{\mu \nu }^{(m)}+E_{\mu \nu },  \label{32}
\end{equation}
where 
\begin{equation}
T_{\mu \nu }^{(n)}=\mu \left( u,r\right) l_{\mu }l_{\nu },
\end{equation}
is the component of the matter field that moves along the null hypersurfaces 
$u=const.$, 
\begin{equation}
T_{\mu \nu }^{(m)}=\left( \rho +p\right) \left( l_{\mu }n_{\nu }+l_{\nu
}n_{\mu }\right) +pg_{\mu \nu },  \label{52}
\end{equation}
represents the energy-momentum tensor of the collapsing matter and 
\begin{equation}
E_{\mu \nu }=\frac{1}{4\pi }\left( F_{\mu \alpha }F_{\nu }^{\alpha }-\frac{1%
}{4}g_{\mu \nu }F_{\alpha \beta }F^{\alpha \beta }\right) ,
\end{equation}
is the electromagnetic contribution. $l_{\mu }$ and $n_{\mu }$ are two
null-vectors given by $l_{\mu }=\delta _{\left( \mu \right) }^{\left(
0\right) }$ and $n_{\mu }=\frac{1}{2}\left[ 1-\frac{2m\left( u,r\right) }{r}%
\right] \delta _{\left( \mu \right) }^{\left( 0\right) }-\delta _{\left( \mu
\right) }^{\left( 1\right) }$, so that $l_{\alpha }l^{\alpha }=n_{\alpha
}n^{\alpha }=0$ and $l_{\alpha }n^{\alpha }=-1$ (with $\delta _{\left(
b\right) }^{\left( a\right) }$ the Kronecker symbol) \cite{Hu96}, \cite
{WaWu98}. The energy density and pressure in Eq. (\ref{52}) have been
obtained by diagonalizing the energy-momentum tensor obtained from the
metric \cite{Hu96}.

The electromagnetic tensor $F_{\mu \nu }$ obeys the Maxwell equations \cite
{LaLi75} 
\begin{equation}
\frac{\partial F_{\mu \nu }}{\partial x^{\lambda }}+\frac{\partial
F_{\lambda \mu }}{\partial x^{\nu }}+\frac{\partial F_{\nu \lambda }}{%
\partial x^{\mu }}=0,  \label{72}
\end{equation}
\begin{equation}
\frac{1}{\sqrt{-g}}\frac{\partial }{\partial x^{\mu }}\left( \sqrt{-g}F^{\mu
\nu }\right) =-4\pi j^{\mu }.  \label{82}
\end{equation}

Without any loss of generality the electromagnetic vector potential can be
chosen as \cite{BoVa70}, \cite{LaZa91} 
\begin{equation}
A_{\mu }=\frac{q\left( u\right) }{r}\delta _{\left( \mu \right) }^{\left(
u\right) },
\end{equation}
with $q(u)$ being an arbitrary integration function. From the Maxwell Eqs. (%
\ref{72}) and (\ref{82}), it follows that the only non-vanishing components
of $F_{\mu \nu }$ are $F_{ru}=-F_{ur}=\frac{q(u)}{r^{2}}$ and, consequently, 
\begin{equation}
E_{\mu }^{\nu }=\frac{q^{2}(u)}{r^{4}}diag\left( -1,1,-1,1\right) .
\end{equation}

For the energy-momentum tensor (\ref{32}) the gravitational field equations
take the form \cite{HaCh00} 
\begin{equation}
\frac{1}{r^{2}}\frac{\partial m\left( u,r\right) }{\partial u}=\frac{1}{2}%
\mu \left( u,r\right) ,  \label{rad}
\end{equation}
\begin{equation}
\frac{2}{r^{2}}\frac{\partial m(u,r)}{\partial r}=\rho \left( u,r\right) +%
\frac{q^{2}(u)}{r^{4}},  \label{dens}
\end{equation}
\begin{equation}
-\frac{1}{r}\frac{\partial ^{2}m\left( u,r\right) }{\partial r^{2}}=p\left(
u,r\right) +\frac{q^{2}(u)}{r^{4}}.  \label{pres}
\end{equation}

The stress-energy tensor (\ref{52}) satisfies the dominant energy condition
if the following three conditions are met: 
\begin{equation}
p\geq 0,\rho \geq p,T_{ab}w^{a}w^{b}>0,
\end{equation}
where $w^{a}$ is an arbitrary time-like four-vector. The first two of these
conditions imply that $\frac{\partial m}{\partial r}\geq 0$ and $\frac{%
\partial ^{2}m}{\partial r^{2}}\leq 0$. The former just says that the mass
function either increases with $r$ or is a constant, which is a natural
physical requirement on it. To satisfy the first two of the dominant energy
conditions one must impose an equation of state for the collapsing matter.

Usually two different equations of state are used for the description of
matter at extremely high densities. One of the most widely investigated case
is the so-called causal limit of the linear barotropic equation of state $%
p=\left( \gamma -1\right) \rho $, $\gamma =$constant, corresponding to $%
\gamma =2$, or the Zeldovich stiff fluid equation of state $p=\rho $ .

The Zeldovich equation of state, valid for densities significantly higher
than nuclear densities, $\rho >10\rho _{nuc}$, with $\rho _{nuc}=10^{14}$
g/cm$^{3}$, can be obtained by constructing a relativistic Lagrangian that
allows bare nucleons to interact attractively via scalar meson exchange and
repulsively via the exchange of a more massive vector meson \cite{ShTe83}.
In the non-relativistic limit both the quantum and classical theories yield
Yukawa-type potentials. At the highest densities the vector meson exchange
dominates and by using a mean field approximation one can show that in the
extreme limit of infinite densities the pressure tends to the energy
density, $p\rightarrow \rho $. In this limit the sound speed $c_{s}=\sqrt{%
dp/d\rho }\rightarrow 1$, and hence this equation of state satisfies the
causality condition, with the speed of sound less than the speed of light 
\cite{ShTe83}.

An alternative approach to the equation of state at ultra-high densities is
based on the assumption that a whole host of baryonic resonant states arise
at high densities. In the Hagedorn model the baryon resonance mass spectrum,
that is the increase in the number of species of particles with mass between 
$m$ and $m+dm$, is given by $dN=N(m)dm\sim a\exp \left( m/T_{H}\right)
/\left( m_{0}^{2}+m^{2}\right) ^{l}dm$, where $N(m)dm$ is the number of
resonances between mass $m$ and $m+dm$. Fitting to the existent experimental
data on baryon resonances show that $m_{0}=500$ MeV and $a=2.63\times 10^{4}$
MeV$^{3/2}$ \cite{Br00}. If $dN$ increased any faster as $m\rightarrow
\infty $ than in the above formula, the partition function would not
converge. Also the partition function converges only if the temperature of
the system is less than $T_{H}$. Thus $T_{H}$, the Hagedorn temperature, is
the effective highest temperature for any system. By interpreting the
transverse momentum distribution of secondaries in very high energy
collisions in terms of the model one obtains $T_{H}\sim 150-190$ MeV and $%
l=5/4$ \cite{Br00}. The corresponding equation of state for the matter is 
\begin{equation}
p=p_{0}+\rho _{0}\ln \frac{\rho }{\rho _{0}},  \label{Hag}
\end{equation}
where $p_{0}=0.314\times 10^{14}$g/cm$^{3}$ and $\rho _{0}=1.253\times
10^{14}$ g/cm$^{3}$ \cite{Ha70}, \cite{Rh71}. The velocity of sound in this
type of matter is $c_{s}=\sqrt{\rho _{0}/\rho }$ . For the Hagedorn equation
of state the speed of sound has the property $c_{s}\rightarrow 0$ for $\rho
/\rho _{0}\rightarrow \infty $, in striking contrast with the mean field
theory approach, leading to the Zeldovich equation of state, in which $%
c_{s}\rightarrow 1$. The Hagedorn equation of state creates new
''particles'' continuously with increasing density, rather than enlarging
the Fermi sea of a single species. The equation of state (\ref{Hag}) could
be valid asymptotically for densities greater than about $10$ times the
nuclear density $\rho _{n}=2\times 10^{14}$ g/cm$^{3}$. The vacuum boundary
of the initial matter distribution is defined by the equation $p=0$,
condition corresponding to a surface density $\rho _{s}=\rho
_{0}e^{-1/4}\approx 0.778\times 10^{14}$ g/cm$^{3}$, two times smaller in
magnitude as the nuclear density. This condition also defines the physical
radius of the initial matter distribution and defines a boundary for the
null fluid.

Hence, as a possible physical model to describe high-density matter in the
final stages of the gravitational collapse we shall adopt the Hagedorn
equation of state. In the following Section we present the general solution
of the gravitational field equations for the null Hagedorn fluid.

\section{Spherical collapse of the Hagedorn null fluid}

From Eq. (\ref{dens}) we immediately obtain 
\begin{equation}
\frac{\rho \left( u,r\right) }{\rho _{0}}=\frac{1}{\rho _{0}r^{2}}\left[ 2%
\frac{\partial m(u,r)}{\partial r}-\frac{q^{2}(u)}{r^{2}}\right] .
\label{dens3}
\end{equation}

With the use of the Hagedorn equation of state Eq. (\ref{Hag}) and of Eq. (%
\ref{dens3}), Eq. (\ref{pres}) can be written in the form 
\begin{equation}
-\frac{1}{r}\left[ \frac{\partial ^{2}m\left( u,r\right) }{\partial r^{2}}+%
\frac{q^{2}(u)}{r^{3}}\right] =p_{0}+\rho _{0}\ln \left\{ \frac{1}{\rho
_{0}r^{2}}\left[ 2\frac{\partial m(u,r)}{\partial r}-\frac{q^{2}(u)}{r^{2}}%
\right] \right\} .  \label{final}
\end{equation}

By introducing a new variable 
\begin{equation}
w\left( u,r\right) =\ln \frac{\rho }{\rho _{0}}=\ln \left\{ \frac{1}{\rho
_{0}r^{2}}\left[ 2\frac{\partial m(u,r)}{\partial r}-\frac{q^{2}(u)}{r^{2}}%
\right] \right\} ,
\end{equation}
it is easy to show that 
\begin{equation}
\frac{1}{\rho _{0}r}\left[ \frac{\partial ^{2}m\left( u,r\right) }{\partial
r^{2}}+\frac{q^{2}(u)}{r^{3}}\right] =e^{w}\left( \frac{1}{2}r\frac{\partial
w}{\partial r}+1\right) .
\end{equation}

Therefore Eq. (\ref{final}) becomes 
\begin{equation}
r\frac{\partial w}{\partial r}=-2\frac{\alpha +w+e^{w}}{e^{w}},  \label{w}
\end{equation}
where $\alpha =p_{0}/\rho _{0}\approx 0.25$. By integrating Eq. (\ref{w}) we
obtain 
\begin{equation}
\frac{r}{C(u)}=\exp \left( -\frac{1}{2}\int \frac{e^{w}}{\alpha +w+e^{w}}%
dw\right) =\exp \left[ -\frac{1}{2}F(w)\right] ,  \label{ratio}
\end{equation}
where $C(u)$ is an arbitrary integration function and we denoted $F(w)=\int 
\frac{e^{w}}{\alpha +w+e^{w}}dw$. Eq. (\ref{ratio}) formally defines $w$ as
a function of $\eta =r/C(u)$, $w=H\left( \eta \right) $. The variation of
the mass function is described by the equation 
\begin{equation}
2\frac{\partial m(u,r)}{\partial r}=\frac{q^{2}(u)}{r^{2}}+\rho
_{0}r^{2}e^{w},
\end{equation}
having the general solution given by 
\begin{equation}
2m(u,r)=D(u)-\frac{q^{2}(u)}{C(u)}e^{\frac{1}{2}F(w)}-\frac{\rho _{0}}{2}%
C^{3}\left( u\right) \int \frac{\exp \left( 2w-\frac{3}{2}F(w)\right) }{%
\alpha +w+e^{w}}dw,  \label{mass}
\end{equation}
where $D\left( u\right) $ is an arbitrary integration function. In the
following we shall also denote 
\begin{equation}
K(w)=\int \frac{\exp \left( 2w-\frac{3}{2}F(w)\right) }{\alpha +w+e^{w}}dw.
\end{equation}

The density and the pressure of the collapsing null Hagedorn fluid are given
by 
\begin{equation}
\rho \left( w\right) =\rho _{0}\exp \left( w\right) ,p\left( w\right)
=p_{0}+\rho _{0}w.  \label{densx}
\end{equation}

The energy density $\mu $ of the radiation moving along the $u=$constant
null hypersurfaces is given by 
\begin{equation}
\mu \left( u,r\right) =\frac{\dot{D}(u)}{C^{2}(u)}e^{F(w)}-\frac{2q(u)\dot{q}%
(u)}{C^{3}(u)}e^{\frac{3}{2}F(w)}-\frac{3\rho _{0}}{2}\dot{C}(u)e^{F(w)}K(w)+%
\frac{\rho _{0}}{2}\dot{C}(u)\frac{e^{2w-F(w)}}{\alpha +w+e^{w}}\frac{dH}{%
d\eta },
\end{equation}
where a dot denotes the derivative with respect to $u$.

The function $r/C(u)$ can be represented as a power series of the parameter $%
w$ in the form 
\begin{equation}
\frac{r}{C(u)}\approx 1-\frac{2w}{5}+\frac{w^{2}}{5}-\frac{w^{3}}{5}+\frac{%
13w^{4}}{60}-\frac{367w^{5}}{1500}+O^{6}[w],
\end{equation}
while the function $K(w)$ has the power series representation 
\begin{equation}
K\left( w\right) \approx \frac{4w}{5}-\frac{8w^{2}}{25}+\frac{52w^{3}}{25}-%
\frac{59w^{4}}{125}+\frac{351w^{5}}{625}+O^{6}[w].
\end{equation}

In the limit of large densities $\rho \rightarrow \infty $, which also
implies $w\rightarrow \infty $, Eq. (\ref{w}) becomes 
\begin{equation}
r\frac{\partial w}{\partial r}\approx -2,
\end{equation}
with the general solution given by 
\begin{equation}
w\left( u,r\right) \approx \ln \frac{C(u)}{r^{2}}.
\end{equation}

Therefore in the asymptotic limit of very high densities, corresponding to $%
r\rightarrow 0,$ the gravitational collapse of the Hagedorn null fluid is
described by the equations 
\begin{equation}
2m\left( u,r\right) \approx D(u)+\rho _{0}C(u)r-\frac{q^{2}(u)}{r},
\label{cent}
\end{equation}
\begin{equation}
\rho \left( u,r\right) \approx \rho _{0}\frac{C(u)}{r^{2}},p\left(
u,r\right) \approx p_{0}+\rho _{0}\ln \frac{C(u)}{r^{2}},
\end{equation}
\begin{equation}
\mu \left( u,r\right) \approx \frac{\dot{D}}{r^{2}}+\rho _{0}\frac{\dot{C}}{r%
}-\frac{2q\dot{q}}{r^{3}}.
\end{equation}

In order to find the behavior of the solution in the opposite limit of
large, but finite $r$, we note first that the boundary of the Hagedorn type
matter distribution is defined by the equation $p=0$, corresponding to a
value of the parameter $w=w_{s}=-p_{0}/\rho _{0}\approx -1/4$. Values of $%
w<w_{s}$ lead to the unphysical situation of negative pressure matter
distribution, $p<0$ for $w<w_{s}$. In order to find the behavior of the
solution near $w_{s}$, we represent $w(u,r)$ in the form $%
w(u,r)=w_{s}+w_{1}(u,r)$, with $w_{1}\left( u,r\right) $ a small
perturbation satisfying the condition $\left| w_{1}\left( u,r\right) \right|
<<\left| w_{s}\right| $, $\forall u,r$. Substituting into Eq. (\ref{w})
gives 
\begin{equation}
r\frac{\partial w_{1}}{\partial r}\approx -2\frac{w_{1}+e^{w_{s}}e^{w_{1}}}{%
e^{w_{s}}e^{w_{1}}}\approx -2\frac{\left( 1+s\right) w_{1}+s}{s\left(
1+w_{1}\right) },  \label{w1}
\end{equation}
where we denoted $s=\exp \left( w_{s}\right) \approx \exp (-1/4) \approx
0.778$. Hence for large $r$ the solution of Eq. (\ref{w1}) is given by 
\begin{equation}
\frac{s}{1+s}w_{1}+\frac{s\ln s}{\left( 1+s\right) ^{2}}+\frac{s}{\left(
1+s\right) ^{2}}\ln \left[ 1+\frac{1+s}{s}w_{1}\right] =-2\ln r+C^{\prime
}(u),  \label{w2}
\end{equation}
where $C^{\prime }(u)$ is an arbitrary integration function. Since we have
assumed that $w_{1}\left( u,r\right) $ is small, we can neglect in Eq. (\ref
{w2}) the term containing the logarithmic function. Hence near the vacuum
boundary of the high density Hagedorn fluid distribution we obtain 
\begin{equation}
w\approx w_{0}+\ln \frac{C(u)}{r^{2k}},
\end{equation}
where $w_{0}=w_{s}s/(1+s)\approx -0.10$, $k=\left( 1+s\right) /s=2.284$ and $%
C(u)$ is an arbitrary, $u$-dependent integration function.

Therefore in the limit $r\rightarrow \infty $ the general solution of the
gravitational field equations can be approximated by 
\begin{eqnarray}
2m\left( u,r\right) &\approx &D(u)-\frac{q^{2}(u)}{r}+\frac{\rho
_{0}e^{w_{0}}}{3-2k}C(u)r^{3-2k}\approx  \nonumber \\
&&D(u)-\frac{q^{2}(u)}{r}-0.6\rho _{0}\frac{C(u)}{r^{1.56}},
\end{eqnarray}
\begin{equation}
\rho \left( u,r\right) \approx \rho _{0}e^{w_{0}}\frac{C(u)}{r^{2k}}\approx
0.9\rho _{0}\frac{C(u)}{r^{4.56}},
\end{equation}
\begin{equation}
p\left( u,r\right) \approx p_{0}+\rho _{0}w_{0}+\rho _{0}\ln \frac{C(u)}{%
r^{2k}}\approx \rho _{0}\left( 0.15+\ln \frac{C(u)}{r^{4.56}}\right) ,
\end{equation}
\begin{eqnarray}
\mu \left( u,r\right) &\approx &2\frac{\dot{D}(u)}{r^{2}}-\frac{2q(u)\dot{q}%
(u)}{r^{3}}+\frac{2\rho _{0}e^{w_{0}}}{3-2k}\dot{C}(u)r^{1-2k}\approx 
\nonumber \\
&&2\frac{\dot{D}(u)}{r^{2}}-\frac{2q(u)\dot{q}(u)}{r^{3}}-0.58\rho _{0}\frac{%
\dot{C}(u)}{r^{3.56}},
\end{eqnarray}
where $D(u)$ is an arbitrary integration function.

The variation of the ratio $r/C(u)$ as a function of $w$ is represented, by
using Eq. (\ref{ratio}), in Fig. 1.

\begin{figure}[h]
\epsfxsize=10cm
\centerline{\epsffile{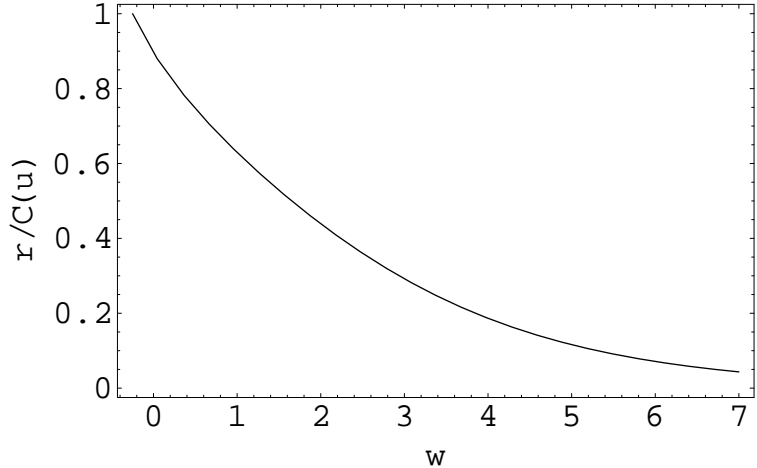}}
\caption{Variation of the ratio $r/C(u)$ as a function of the parameter $w$.}
\label{FIG1}
\end{figure}

As one can see from the figure, in the limit $w\rightarrow w_{s}=-0.25$, the
ratio $r/C(u)$ tends to $1$, while for large $w$, $w\rightarrow \infty $,
that is, in the limit of very high densities, $r/C(u)$ tends to zero. Values
of $w<w_{s}=-0.25$ lead to the violation of the dominant energy condition $p\geq 0$%
. The function $K(w)$ is represented, as a function of $r/C(u)$, in Fig. 2.

\begin{figure}[h]
\epsfxsize=10cm
\centerline{\epsffile{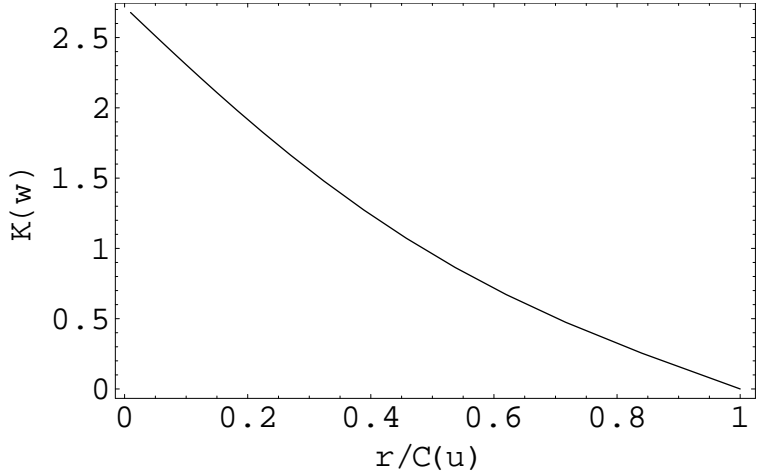}}
\caption{Variation of $K(w)$ as a function of $r/C(u)$.}
\label{FIG2}
\end{figure}

For the sake of comparison we shall also present the solution of the
gravitational field equations corresponding to the collapse of the charged
null Zeldovich fluid with $\rho =p$. In this case Eqs. (\ref{dens}) and (\ref
{pres}) give immediately 
\begin{equation}
\frac{2}{r^{2}}\frac{\partial m\left( u,r\right) }{\partial r}+\frac{1}{r}%
\frac{\partial ^{2}m\left( u,r\right) }{\partial r^{2}}=0,
\end{equation}
with the general solution given by 
\begin{equation}
m(u,r)=B(u)-\frac{A^{2}(u)}{r},
\end{equation}
where $A(u)$ and $B(u)$ are arbitrary integration functions.

The other physical quantities characterizing the collapsing Zeldovich fluid
are given by 
\begin{equation}
\rho (u,r)=p(u,r)=\frac{2A^{2}(u)-q^{2}(u)}{r^{4}},
\end{equation}
\begin{equation}
\mu \left( u,r\right) =\frac{2}{r^{2}}\left( \frac{dB}{du}-\frac{2}{r}A(u)%
\frac{dA}{du}\right) .
\end{equation}

In all these cases the electromagnetic current follows from the Maxwell Eq. (%
\ref{82}) and can be generally represented as 
\begin{equation}
j^{\mu }=\frac{1}{4\pi r^{2}}\frac{dq(u)}{du}l^{\mu }.
\end{equation}

The energy-momentum tensor of the mixture of fluids under consideration
belongs to the Type II fluids \cite{Hu96}. The energy conditions are the
weak, strong and dominant energy conditions $\mu \geq 0,\rho \geq 0,p\geq
0,\rho \geq p\geq 0$ and can be satisfied for both Hagedorn and Zeldovich
fluids by appropriately choosing the arbitrary functions $A(u)$, $B(u)$, $%
C(u)$ and $D(u)$ that characterize the injection and initial distribution of
the mass and $q(u)$ that describes the variation of the charge. For the
Hagedorn fluid the conditions $\rho \geq 0$ and $p\geq 0$ are trivially
satisfied for the values of the parameter $w$ in the range $w\in \left(
w_{s},\infty \right) $, corresponding to the range of densities of $\rho \in
\left( \rho _{s},\infty \right) $, with $w_{s}=-0.25$ and $\rho
_{s}=0.78\times 10^{14}$ g/cm$^{3}$. Due to the choice of the parameters in
the equation of state the dominant energy condition $\rho \geq p$ is also
satisfied. The condition $\mu \geq 0$ is equivalent to $\frac{dm}{du}\geq 0$%
, and in the approximation of high densities leads to 
\begin{equation}
\frac{dD(u)}{du}+\rho _{0}\frac{dC(u)}{du}r\geq 2q(u)\frac{dq(u)}{du}\frac{1%
}{r},  \label{22}
\end{equation}
imposing a simultaneous constraint on all three functions $C(u),D(u)$ and $%
q(u)$. For small values of $r$ and for a charged collapsing high density
fluid, the right hand side of Eq. (\ref{22}) dominates and this energy
condition could not hold. One possibility to satisfy Eq. (\ref{22}) for all $%
r$ is to assume that the function $q(u)$ behaves so that $\frac{dq^{2}(u)}{du%
}\rightarrow 0$ for $r\rightarrow 0$. This means that the charge in the
singular point $r=0$ is constant for all times. Alternatively, we may
suppose that at extremely small radii matter is generated so as to satisfy
the energy condition. For neutral $q\equiv 0$ matter Eq. (\ref{22}) is
easily satisfied by choosing $\frac{dC(u)}{du}>0$ and $\frac{dD(u)}{du}>0$.
In the case of the Zeldovich fluid the energy conditions are satisfied by
choosing the functions $A$, $B$ and $q$ so that $A^{2}(u)\geq q^{2}(u)/2$
and $\dot{B}(u)\geq 2A(u)\dot{A}\left( u\right) /r$.

The radii of the apparent horizon of the metric (\ref{222}) are given by the
solution of the equation $2m=r$. If $\lim_{u\rightarrow \infty
}C(u)=C_{0}=const.$, $\lim_{u\rightarrow \infty }D(u)=D_{0}=const.$ and $%
\lim_{u\rightarrow \infty }q(u)=q_{0}=const.$, then the algebraic equation
determining the radii of the apparent horizons in the case of the Hagedorn
fluid is 
\begin{equation}
C_{0}e^{-\frac{1}{2}F(w)}+\frac{\rho _{0}}{2}C_{0}^{3}K(w)+\frac{q_{0}^{2}}{%
C_{0}}e^{\frac{1}{2}F(w)}=D_{0},
\end{equation}
which in general may have multiple solutions.

In the case of the collapse of the Zeldovich fluid the radii of the apparent
horizon are given by the solutions of the equation 
\begin{equation}
2B_{0}-2\frac{A_{0}^{2}}{r}=r,
\end{equation}
where $A_{0}=\lim_{u\rightarrow \infty }C(u)=$constant and $%
B_{0}=\lim_{u\rightarrow \infty }D(u)=$constant. The radii of the apparent
horizon are 
\begin{equation}
r_{1,2}=B_{0}\pm \sqrt{B_{0}^{2}-2A_{0}^{2}}.
\end{equation}

The singularities of the matter filled Vaidya space-time can be recognized
from the behavior of the energy density and curvature scalars like e.g. $%
R_{\alpha \beta }R^{\alpha \beta }$ and $R_{\alpha \beta \gamma \delta
}R^{\alpha \beta \gamma \delta }$, given in the high density limit for the
collapsing Hagedorn fluid by 
\begin{equation}
R_{\alpha \beta }R^{\alpha \beta }=\frac{2}{r^{8}}\left[ r^{4}\rho
_{0}^{2}C^{2}(u)+2r^{2}\rho _{0}C(u)q^{2}(u)+2q^{4}(u)\right] ,
\end{equation}
\begin{equation}
R_{\alpha \beta \gamma \delta }R^{\alpha \beta \gamma \delta }=4\frac{%
r^{4}\rho _{0}^{2}C^{2}(u)-2r^{3}\rho _{0}C(u)D(u)+3r^{2}D^{2}(u)-2r^{2}\rho
_{0}C(u)q^{2}(u)+12rD(u)q^{2}(u)+14q^{4}(u)}{r^{8}},
\end{equation}
and which diverge for $r\rightarrow 0$.

\section{Outgoing radial null geodesics equation}

The central shell-focussing singularity (i.e. that occurring at $r=0$) is
naked if the radial null-geodesic equation admits one or more positive real
roots $X_{0}$ \cite{Jo93}. In the case of the pure Vaidya space-time it has
been shown that for a linear mass function $2m(u)=\lambda u$ the singularity
at $r=0$, $u=0$ is naked for $\lambda \leq \frac{1}{8}$ \cite{DwJo91}. Hence
it is important to investigate whether the gravitational collapse of high
density matter described by the Hagedorn and Zeldovich equations of state
could result in the formation of naked singularities.

We consider first the case of the gravitational collapse of the Hagedorn
fluid. In order to simplify the calculations we chose some simple particular
expressions for the functions $C(u),D(u)$ and $q(u)$, e.g. $C(u)=\alpha
_{0}u $, $D(u)=\beta _{0}u$ and $q(u)=q_{0}u$, \ with $\alpha _{0}>0,\beta
_{0}>0$ and $q_{0}\geq 0$ constants. With this choice the equation of the
radially outgoing, future-directed null geodesic originating at the
singularity can be written as 
\begin{equation}
\frac{du}{dr}=\frac{2}{1-\beta _{0}\left( \frac{u}{r}\right)
+q_{0}^{2}\left( \frac{u}{r}\right) ^{2}+\frac{\rho _{0}\alpha _{0}^{3}}{2}%
u^{2}\left( \frac{u}{r}\right) K(w)}.  \label{24}
\end{equation}

For the geodesic tangent to be uniquely defined and to exist at the singular
point $r=0,u=0$ of Eq. (\ref{24}) the following condition must hold \cite
{Jo93}: 
\begin{equation}  \label{25}
\lim_{u,r\rightarrow 0}\frac{u}{r}=\lim_{u,r\rightarrow 0}\frac{du}{dr}%
=X_{0}.
\end{equation}

When the limit exists and $X_{0}$ is real and positive, there is a future
directed, non-space-like geodesic originating from $r=0,u=0$. In this case
the singularity will be, at least, locally naked.

Since the function $K(w)$ is finite for all $w=H\left( r/C(u)\right) $, and
we assume that $\lim_{u,r\rightarrow 0}u/r$ is also finite, it follows that $%
\lim_{u,r\rightarrow 0}u^{2}\left( \frac{u}{r}\right) K(w)=0$. Therefore it
follows that for the null geodesic Eq. (\ref{24}) condition (\ref{25}) leads
to the following third order algebraic equation: 
\begin{equation}
q_{0}^{2}X_{0}^{3}-\beta _{0}X_{0}^{2}+X_{0}-2=0.  \label{26}
\end{equation}

In the case of a neutral Hagedorn fluid with $q(u)\equiv 0$, Eq. (\ref{26})
reduces to a second order algebraic equation with two roots, $%
X_{01,2}=\left( 1\pm \sqrt{1-8\beta _{0}}\right) /2\beta _{0}$. Therefore
the condition for the formation of a naked singularity is $\beta _{0}<1/8$.
For $\beta _{0}>1/8$, as a result of the collapse of the Hagedorn type
fluid, a black hole is formed.

For $q(u)\neq 0$ the condition of the existence of at least one real
solution of Eq. (\ref{26}) is 
\begin{equation}
108q_{0}^{4}+4q_{0}^{2}+8\beta _{0}^{3}\geq \beta _{0}^{2}+36\beta
_{0}q_{0}^{2}.
\end{equation}

Therefore, by appropriately choosing the constants $q_{0}$ and $\beta _{0}$,
it is always possible to construct a positive solution of Eq. (\ref{26}).

In the case of the collapse of the Zeldovich fluid and by assuming for the
arbitrary $u$-dependent integration functions the form $A(u)=a_{0}u$ and $%
B(u)=b_{0}u$, with $a_{0},b_{0}$ non-negative constants, the equation of the
radially outgoing, future-directed null geodesic originating at the
singularity is 
\begin{equation}
\frac{du}{dr}=\frac{2}{1-b_{0}\left( \frac{u}{r}\right) +a_{0}^{2}\left( 
\frac{u}{r}\right) ^{2}}.
\end{equation}

The algebraic condition for the formation of a naked singularity is given by
the requirement that the equation 
\begin{equation}
a_{0}^{2}X_{0}^{3}-b_{0}X_{0}^{2}+X_{0}-2=0,  \label{Zeld}
\end{equation}
has at least one positive root $X_{0}>0$. The condition that the above
equation has at least one real root can be written as 
\begin{equation}
\frac{a_{0}^{2}}{b_{0}^{2}}\geq \frac{1-8b_{0}}{108a_{0}^{2}-36b_{0}+4}.
\end{equation}

This condition can be satisfied, for example, by choosing $b_{0}<1/8$ and $%
a_{0}>1/\sqrt{216}$. Therefore, as in the case of the charged Hagedorn
fluid, it is always possible to construct a positive solution of Eq. (\ref
{Zeld}).

\section{Collapsing Hagedorn matter-a possible source of Gamma-ray bursts}

Gamma-ray bursts (GRBs) are cosmic gamma ray emissions with typical fluxes
of the order of $10^{-5}$ to $5\times 10^{-4}$erg cm$^{-2}$ with the rise
time as low as $10^{-4}$ s and the duration of bursts from $10^{-2}$ to $%
10^{3}$ s \cite{Pi98}. The distribution of the bursts is isotropic and they
are believed to have a cosmological origin, recent observations suggesting
that GRBs might originate at extra-galactic distances \cite{Pi98}. The large
inferred distances imply isotropic energy losses as large as $3\times
10^{53} $ erg for GRB 971214 and $3.4\times 10^{54}$ erg for GRB 990123 \cite
{Ku99}. In the present Section we shall use CGS units.

The widely accepted interpretation of the phenomenology of $\gamma $-ray
bursts is that the observable effects are due to the dissipation of the
kinetic energy of a relativistically expanding fireball, whose primal cause
is not yet known \cite{Pi98}.

The proposed models for the energy source involve merger of binary neutron
stars \cite{Pi92}, capture of neutron stars by black holes \cite{Ca92},
differentially rotating neutron stars \cite{KlRu98} or neutron star-quark
star conversion \cite{ChDa96} etc. However, the most popular model involves
the violent formation of an approximately one solar mass black hole,
surrounded by a similarly massive debris torus. The formation of the black
hole and debris torus may take place through the coalescence of a compact
binary or the collapse of a quickly rotating massive stellar core \cite{MeRe93}. There
are still many open problems concerning GRBs, from which the most important
is the problem of the source of the large energy emission during the bursts.
On the other hand naked singularities as sources of $\gamma $-ray bursts
have also been proposed \cite{ChJo94}, \cite{Si98}, \cite{HaCh00a}. The fact
that explosive radiation can be emitted during the gravitational collapse to
a naked singularity of a dust ball has also been pointed out recently by 
\cite{HaIgNa00}.

As an astrophysical application of the Hagedorn fluid collapse in the Vaidya
geometry we consider the possibility that gamma - ray bursts could in fact
be energy emission during the collapse of a neutral, $q\equiv 0$, high
density matter in the Hagedorn phase, ending with the formation of a naked
singularity. An estimation of the energy emitted during the collapse shows
that it is of the same order of magnitude as the one measured during $\gamma
-$ ray bursts. Hence this mechanism could provide a valuable explanation for
this phenomenon, also opening the possibility of the observational
investigation of the astrophysical properties of the naked singularities.

The arbitrary integration functions $C(u)$ and $D(u)$ appearing in Eq. (\ref
{mass}), describe the injection and initial distribution of the mass,
respectively. Their exact mathematical form cannot be obtained in the
framework of the present formalism. The only requirements that the functional
form of these functions must satisfy are the dominant energy conditions
and the condition that the equation of the radially
outgoing, future-directed null geodesics, originating at the singularity $%
r=0 $, be uniquely defined and exist at the singular point $r=0$ and has at
least one positive root. A positive root of the geodesic equation leads to
the possibility of formation of a naked singularity as a result of the
collapse. Some simple integration functions satisfying the condition on the
geodesic equation, as well as the dominant energy conditions, are, for example, $C(u)=\alpha _{0}u$, $D(u)=\beta _{0}u$,
with $\alpha _{0}>0$, $\beta _{0}>0$ constants. With this choice a naked
singularity may form during the collapse of the Hagedorn fluid if and only
if the algebraic equation $\beta _{0}X_{0}^{2}-X_{0}+2=0$ has at least one
positive root $X_{0}>0$. This condition requires $\beta _{0}<1/8$. Of course
many other choices are also possible.

Naked singularities are gravitational singularities that are not covered by
a horizon. Near the singularity the space-time curvature and the
gravitational tidal forces grow very strongly. During the collapse naked
singularities could emit powerful bursts of radiation visible to a distant
external observer situated far away from the sight of the collapse \cite
{ChJo94}.

As a first physical parameter we need to estimate is the time $t_{ff}$
necessary for a matter element at the surface of the star to reach the
center $r=0$ of the collapsed object, as a function of the mass at the
center. This can be done by evaluating the time derivative of the mass,
given by Eq. (\ref{cent}) with $q=0$, at the center of the naked
singularity: 
\begin{equation}
\frac{dm}{dt}\mid _{r=0}=\frac{c^{3}}{2G}\left( \frac{dD(u)}{du}\right)
_{r=0}\text{.}  \label{mt}
\end{equation}

By integrating the above equation from $t=0$ to $t_{ff}$ we obtain 
\begin{equation}
M\mid _{r=0}=\frac{c^{3}}{2G}\int_{0}^{t_{ff}}\left( \frac{dD(u)}{du}\right)
_{r=0}dt,  \label{wab}
\end{equation}
where we assumed $m(0)\mid _{r=0}=0$ and denoted $m(t_{ff})\mid _{r=0}=M\mid
_{r=0}=0$.
In order to find an explicit expression for $t_{ff}$ we shall approximate
the integral in Eq. (\ref{wab}) by using the first mean value theorem for
integrals, which states that for any arbitrary function $f(t)$, $%
\int_{a}^{b}f(t)dt=(b-a)f(c)$, where $a<c<b$. $f(c)$ is called the average
value of $f$. Hence, in the following we shall approximate the derivatives of the arbitrary
the arbitrary integration functions and the functions themselves by their average values.
Therefore from Eq. (\ref{wab}) we obtain 
\begin{equation}
t_{ff}=\frac{2G}{c^{3}\beta _{0}}M\mid _{r=0},  \label{time}
\end{equation} where we denoted by $\beta _{0}\geq 0$ the average value of the function $%
\left( dD(u)/du\right) _{r=0}$,
$\beta _{0}=\left[ dD(t)/d(t)\right] _{t=\theta }$, where $0<\theta <t_{eff}$.

The initial mass distribution of the Hagedorn fluid can be obtained from the
study of the mass distribution at the initial moment $t=0$: 
\begin{equation}
\frac{dm}{dr}\mid _{t=0}=\frac{c^{2}}{2G}\left[ \frac{dD(u)}{du}\mid
_{t=0}+\rho _{0}r\frac{dC(u)}{du}\mid _{t=0}\right] .
\end{equation}

By integrating this equation, and using again the first mean value theorem, it
follows that the initial mass profile of the Hagedorn fluid can be
represented as 
\begin{equation}
m(r)\mid _{t=0}=\frac{c^{2}}{2G}\alpha _{0}r,  \label{prof}
\end{equation}
where $\alpha _{0}\geq 0$ is the average value of the $r$-dependent function 
$h(r)=\left[ dD(u)/du+\rho _{0}r(dC(u)/du)\right] \mid _{t=0}$, with the
function $h(r)$ estimated at some point $r=\sigma $, $0<\sigma <R$, with $R$ the
radius of the Hagedorn fluid distribution. Hence $\alpha _{0}=h(\sigma )$. From this
equation we can see that our approach implies a linear profile of the
initial mass distribution of the star. $\alpha _{0}$, the average value of
the spatial derivatives of the integration functions $D(u)$ and $C(u)$ at $%
t=0$, is completely determined by the initial conditions, that is by the
total initial mass of the star $M\mid _{t=0}$ and by its initial radius $R$
via the relation 
\begin{equation}
\alpha _{0}=\frac{2GM\mid _{t=0}}{c^{2}R}.
\end{equation}

In order to calculate the energy emitted during the Hagedorn fluid collapse
we shall admit that the luminosity of the collapsing object should not
exceed the rate of collapsing matter energy \cite{HaCh00a}. The variation of
the mass of the high density fluid during collapse is given by 
\begin{equation}
\Delta \frac{2Gm(u,r)}{c^{2}}=\left[ \Delta D(u)+\rho _{0}r\Delta C(u)+\rho
_{0}C(u)\Delta r\right] \approx \left[ \frac{dD(u)}{du}\Delta u+\rho _{0}r%
\frac{dC(u)}{du}\Delta u+\rho _{0}C(u)\Delta r\right] .
\end{equation}

The variation of the advanced time coordinate $u$ is given by the
approximate expression $\Delta u=c\Delta t+\Delta r=c\Delta t\left( 1+\frac{1%
}{c}\frac{\Delta r}{\Delta t}\right) \approx c\Delta t\left( 1+\frac{v_{f}}{c%
}\right) $, where we defined $v_{f}=\frac{\Delta r}{\Delta t}$ as being the
speed of the collapsing fluid as measured by a local observer. Then we
obtain 
\begin{equation}
\Delta \frac{2Gm(u,r)}{c^{2}}\approx \left[ \frac{dD(u)}{du}+\rho _{0}r\frac{%
dC(u)}{du}+\rho _{0}C(u)\frac{v_{f}}{c}\left( 1+\frac{v_{f}}{c}\right) ^{-1}%
\right] c\left( 1+\frac{v_{f}}{c}\right) \Delta t.  \label{en1}
\end{equation}

We assume that the energy emission occurs mainly from a small region near
the center of the naked singularity. Hence we shall evaluate Eq. (\ref{en1})
near $r$ very close to zero. Then from Eq. (\ref{mt}) we can roughly
approximate $dD(u)/du\approx \left( 2G/c^{3}\right) (dm/dt)\approx
2G/c^{3}\left( M/t_{ff}\right) =\beta _{0}$. Since for a small $r$, $C\left(
ct+r\right) \approx C\left( ct\right) $, we can also neglect in this limit
the term $\rho _{0}r\left( dC(u)/du\right) $. From the field Eq. (\ref{dens}%
) and from Eq. (\ref{densx}) it immediately follows that $2\left( \partial
m\left( u,r\right) /\partial r\right) =\left( 8\pi G/c^{2}\right) r^{2}\rho
\approx \rho _{0}C(u)$. Taking into account Eq. (\ref{prof}) we immediately
find $\rho _{0}C(u)\approx 4\pi \alpha _{0}\approx \left( 8\pi
GM/c^{2}R\right) $.
The velocity $v_{f}$ of the collapsing matter can be obtained from $%
v_{f}\approx R/t_{ff}\approx (c^{3}R\beta _{0})/\left( 2GM\mid _{r=0}\right) 
$. 

With the use of the previous results we obtain for the total energy emitted
per unit time during the collapse of a Hagedorn fluid to a naked singularity the
expression: 
\begin{equation}
\frac{\Delta E_{r}}{\Delta t}\approx \frac{1}{2}\left[ \beta _{0}+4\pi
\alpha _{0}\frac{v_{f}}{c}\left( 1+\frac{v_{f}}{c}\right) ^{-1}\right]
\left( 1+\frac{v_{f}}{c}\right) \frac{c^{5}}{G}\text{ erg s}^{-1}.
\label{en}
\end{equation}

Hence we have expressed the total energy of the radiation, which could be
emitted from a naked singularity, in terms of the average values of the
derivatives of the arbitrary integration functions. By assuming an emission
time of the order of $\Delta t\sim 10^{-4}$ s and a velocity of the
collapsing matter of the order $v_{f}=6\times 10^{8}$cm/s we obtain a value
of the emitted energy $\Delta E_{\gamma }\approx 1.85\times \left[ \beta
_{0}+4\pi \alpha _{0}\frac{v_{f}}{c}\left( 1+\frac{v_{f}}{c}\right) ^{-1}%
\right] \times 10^{55}$erg. Therefore the energy which could be emitted during
the collapse of the Hagedorn fluid to a naked singularity could exceed in
magnitude the energy $\Delta E_{\gamma }\approx 4\times 10^{54}$ erg,
observed for GRB990123 \cite{Ku99}. Of course the exact value of the energy
depends on the exact numerical values of the constants $\alpha _{0}$ and $%
\beta _{0}$. By comparing Eq. (\ref{en}) with the observational data it is
possible to determine the values of $\alpha _{0}$ and $\beta _{0}$. 

Under the effect of the collapse the compact object will heat up to a
temperature of the order of $T\sim T_{H}\sim 10^{13}$K, higher than that
occurring in super nova explosions. Since $T_{H}$ cannot be exceeded, the
gravitational energy of the compact object is converted into new particles,
which will be generated during the collapse. Most of the newly created
particle will decay via weak interaction processes. As a result a
neutrinosphere will form around the naked singularity. Therefore the main
energy loss mechanism of the super-heated collapsed astrophysical object would be
neutrino radiation. The neutrinos and anti neutrinos interact with protons
and neutrons via the URCA processes $n+\nu _{e}\rightarrow p+e^{-}$, $p+\bar{%
\nu}_{e}\rightarrow n+e^{+}$. At temperatures higher than the nuclear Fermi
temperature $kT^{F,N}=\left( \frac{6\pi ^{2}}{g}\right) ^{2/3}\left( \frac{%
\hbar }{2m}\right) \left( \frac{N}{V}\right) ^{2/3},$ that can also be
expressed in the form $T_{11}^{F,N}=1.47\times 10^{-3}\rho _{9}^{2/3}$ \cite
{MeRe93}, the integrated optical neutrino depth is unity. Hence the
deposition energy can be estimated as $E\approx \left( 1-e^{-\tau }\right)
\Delta E_{\gamma }$ ergs. The process $\gamma +\gamma \rightarrow
e^{+}+e^{-} $ will generate a fireball that will expand outward. The
expanding shell interacts with the inter stellar medium surrounding the
Hagedorn type naked singularity, and the kinetic energy is finally radiated
through non-thermal processes in shocks \cite{ChDa96}.

The energy released during the collapse of the Hagedorn and radiation fluids
into a naked singularity has the same order of magnitude as that observed in
the case of gamma ray bursts. This strongly suggests the possibility that
gamma ray bursts could be massive compact objects, formed from a Hagedorn
fluid, collapsing to a naked singularity in a cosmological environment.

\section{Conclusions}

In the present paper we have considered the collapse of a Hagedorn type
fluid in the Vaidya geometry.

The exact solution obtained represents the generalization, for Hagedorn type
matter, of the collapsing solutions previously obtained by Vaidya \cite{Va51}%
, Bonnor and Vaidya \cite{BoVa70}, Lake and Zannias \cite{LaZa91} and Husain 
\cite{Hu96}. From a mathematical point of view the solution is represented
in a parametric form. It satisfies all the energy conditions and
consequently describes the collapse to a singular state. The possible
occurrence of a central naked singularity has also been investigated and it
has been shown that, at least for a particular choice of the parameters, a
naked singularity is formed. Depending on the initial distribution of
density and velocity and on the constitutive nature of the collapsing
matter, either a black hole or a naked singularity is formed. The values of
the parameters in the solution (\ref{mass}) and (\ref{densx}) determine
which of these possibilities occurs. The solution describing the collapse of
both the Hagedorn and Zeldovich matter is asymptotically flat, but this
condition does not play any significant role in the formation of the naked
singularity.

The Hagedorn type matter may reside as a permanent component of neutron
stars core at high temperatures or densities and form stable compact stellar
objects. In fact, from a physical point of view, it seems that the high
density limit for the equation of state described by the Hagedorn equation
of state is one of the best and more realistic candidates for the study of
properties of collapsing objects. It also serves to illustrate the much
richer interplay that can occur among particle physics and general
relativity, when more involved quantum field theoretical models are
considered.

As a possible simple astrophysical application of this collapsing solution
we have considered the possibility that gamma ray bursts could be energy
emission during the collapse of a high density star, ending in the formation
of a naked singularity. The radiated energy during this process could be as
high as $10^{55}$ergs. Thus the naked singularity explosion could be a
candidate for the central engine of a gamma ray burst.

\end{document}